\documentclass[a4paper]{jpconf}
\usepackage{graphicx}
\usepackage{cite}

\begin{document}
\title{D meson-hadron angular correlations in pp and p-Pb collisions with ALICE at the LHC}

\author{Fabio Colamaria, for the ALICE collaboration}

\address{Dipartimento Interateneo di Fisica ‘M. Merlin’ and Sezione INFN, Bari, Italy}

\ead{fabio.colamaria@ba.infn.it}

\begin{abstract}
The comparison of angular correlations between charmed mesons and charged hadrons produced in pp, p-Pb and Pb-Pb collisions can give insight into charm quark energy loss mechanisms in hot nuclear medium formed in heavy-ion collisions and can help to spot possible modifications of charm quark hadronization induced by the presence of the medium. The analysis of pp and p-Pb data and the comparison with predictions from pQCD calculations, besides constituting the necessary baseline for the interpretation of Pb-Pb results, can provide relevant information on charm production and fragmentation processes.

We present a study of azimuthal correlations between D$^0$ and D$^{\ast +}$ mesons and charged hadrons measured by the ALICE experiment in pp collisions at $\sqrt{s} = 7$ TeV and p-Pb collisions at $\sqrt{s_{\rm NN}} = 5.02$ TeV. D mesons were reconstructed from their hadronic decays at central rapidity and in the transverse momentum range $2 < p_{_{\rm T}} < 16$ GeV/$c$, and they were correlated to charged hadrons reconstructed in the pseudorapidity range $|\eta| < 0.8$.
\end{abstract}

\section{Physics motivations}
ALICE~\cite{bib:ALICE} measured $p_{_{\rm T}}$-differential cross sections for D-meson production at central rapidity in pp~\cite{bib:Dmes_pp_7, bib:Ds_pp_7, bib:Dmes_pp_2.76} and Pb-Pb~\cite{bib:Dmes_PbPb} collisions. A significant suppression of D-meson yields was found in central Pb-Pb collisions for $p_{_{\rm T}} >$ 4-5 GeV/$c$, with respect to cross sections in pp collisions, scaled by the nuclear overlap function. This can be attributed, at least partially, to the energy loss of charm quarks in the Quark Gluon Plasma formed in such collisions. The study of angular correlations between D mesons and unidentified charged hadrons provides more insight into this topic, giving the possibility to:
\begin{itemize}
\item study the charm quark fragmentation in the various collision systems;
\item address the production processes of $c\overline{c}$ pairs and their interplay with the underlying event;
\item characterize from a different point of view the charm quark energy loss in heavy-ion collisions compared to cross section measurements~\cite{bib:Dmes_PbPb}.
\end{itemize}

To this end, the pp and p-Pb measurements provide a reference for the comparison of the studied observables, i.e.\ the yields in the near side/away side peaks, the integrated yield, and the height of the correlation baseline.

\section{D-meson reconstruction with ALICE}
ALICE is a general purpose detector, optimized to operate in a high track density environment as that of Pb-Pb collisions. It comprises a central barrel in which tracks with $|\eta| < 0.9$ and $p_{_{\rm T}} > 80$ MeV/$c$ are reconstructed and a forward muon spectrometer covering the rapidity range $2.5 < y < 4$.

The main detectors used in this analysis are:
\begin{itemize}
\item Inner Tracking System (ITS), a silicon detector composed of two layers of pixel (SPD), two of drift (SDD) and two of strip (SSD) detectors, with tracking and vertexing purposes;
\item Time Projection Chamber (TPC), a gaseous detector for 3-dimensional tracking and PID via measurement of the specific particle energy loss;
\item Time Of Flight (TOF), which provides PID, exploited in this analysis for separating pions from kaons up to $2.5$ GeV/$c$ of momentum.
\end{itemize}

D mesons are reconstructed at central rapidity via their hadronic decays by performing an invariant mass analysis of displaced secondary vertices, exploiting the topology of the decay and the secondary vertex displacement together with particle (K and $\pi$) identification.

In particular, D$^0$, D$^+$, D$^{\ast +}$ and D$^+_s$ mesons have been reconstructed in the D$^0$$\rightarrow$$\rm{K}^{-}\pi^+$, D$^+$$\rightarrow$K$^-\pi^+\pi^+$, D$^{\ast +}$$\rightarrow$D$^0\pi^+$, D$^+_s$$\rightarrow$$\phi\pi^+$$\rightarrow$K$^-$K$^+\pi^+$ decay channels.

\begin{figure}[!ht]
\begin{minipage}{0.40\linewidth}\hspace*{0.3cm}
\includegraphics[width=\linewidth]{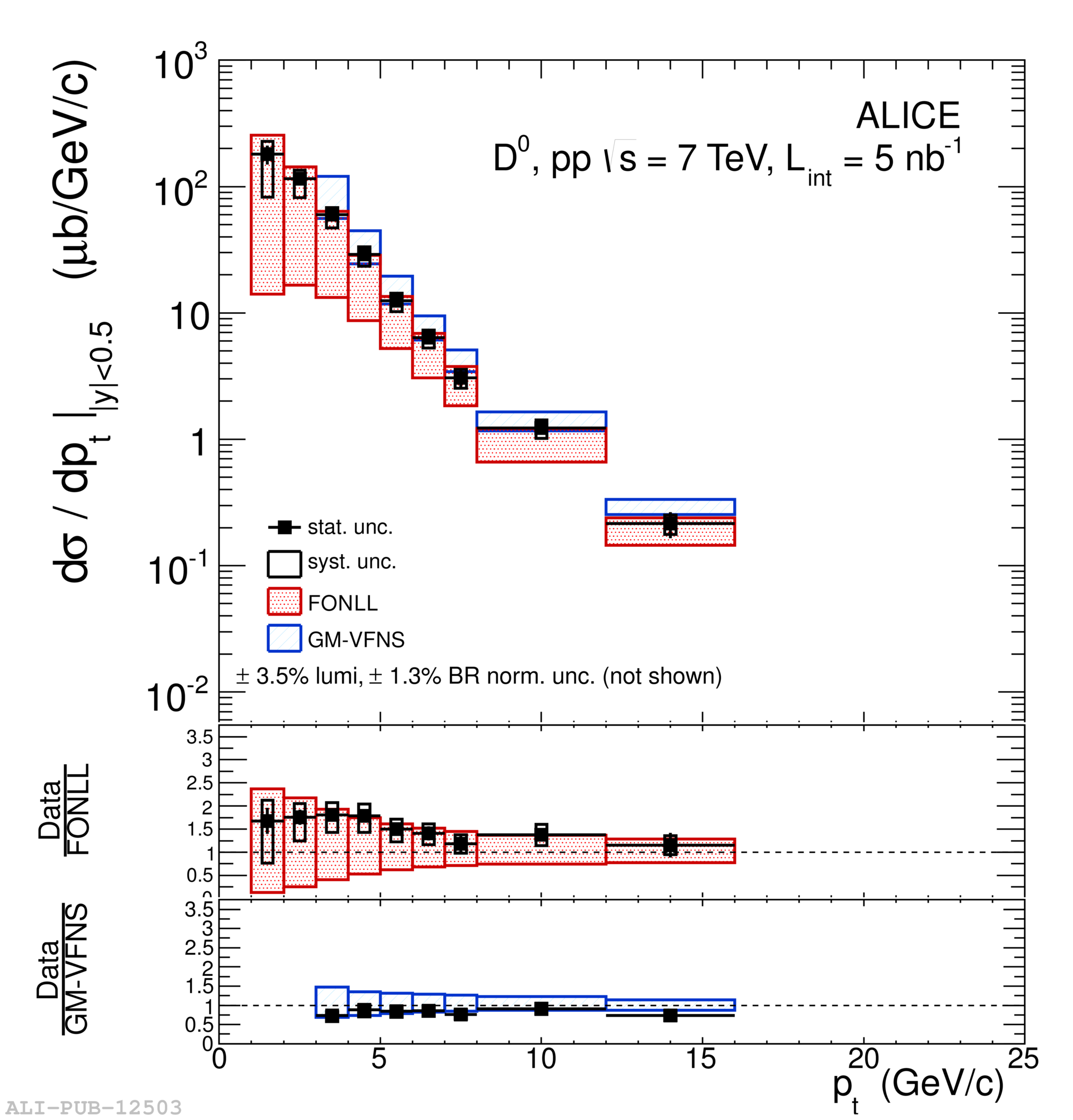}
\end{minipage}\hspace*{0.6cm}
\vspace{-0.5pc}
\begin{minipage}{0.52\linewidth}
\includegraphics[width=\linewidth]{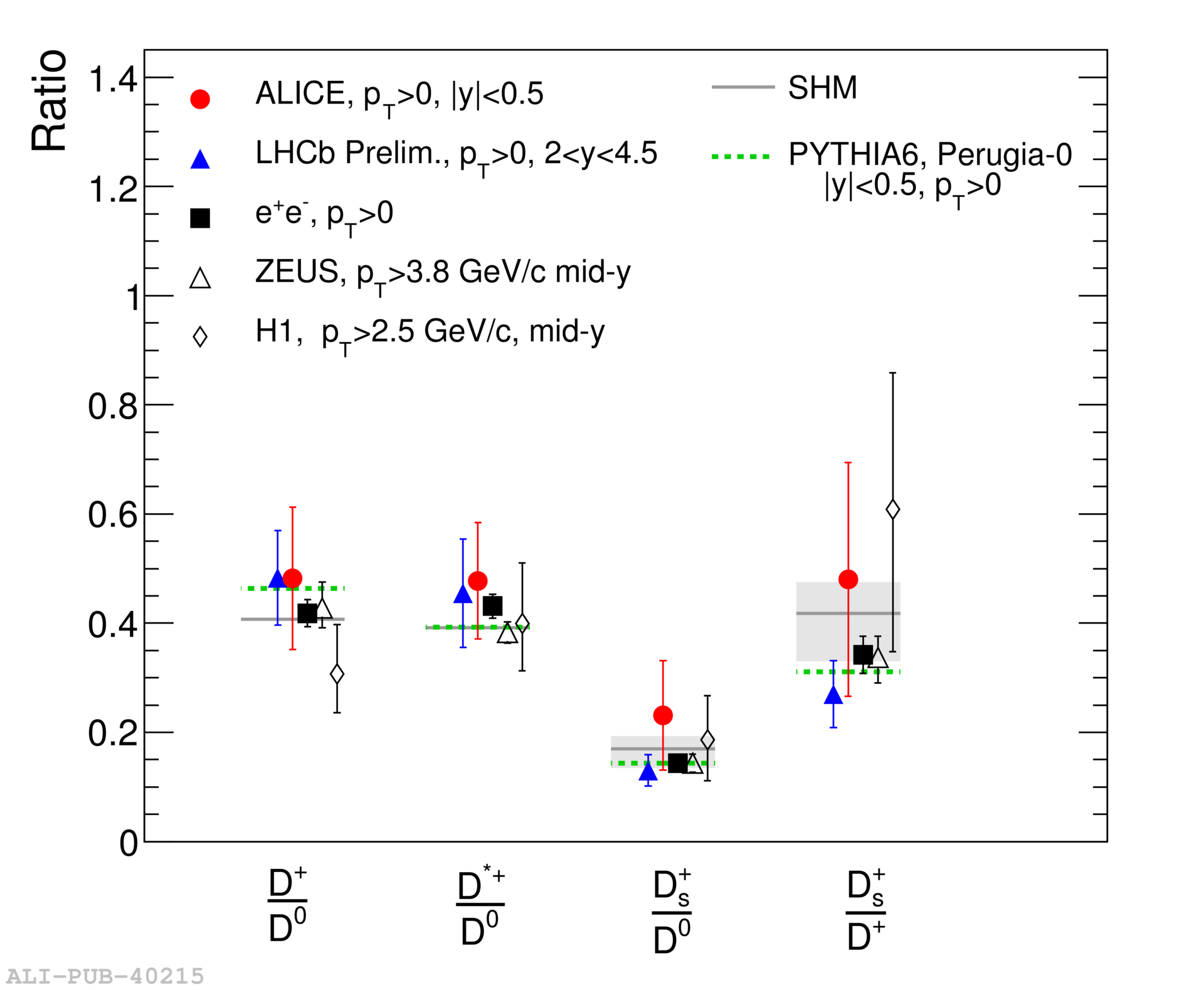}
\end{minipage}
\caption{\label{fig:DCross}Left: cross section for prompt D$^0\ $production at central rapidity in pp collisions at $\sqrt{s} = 7$ TeV, compared to FONLL~\cite{bib:FONLL} (red boxes) and GM-VFNS~\cite{bib:GM-VFNS} (blue boxes) calculations. Ratios between data and calculations are shown in the bottom panels. Right: ratio of cross sections for D$^0$, D$^+$, D$^{\ast +}$ and D$^+_s$ production in pp collisions at $\sqrt{s} = 7$ TeV (red circles). Results from other experiments~\cite{bib:LHCb, bib:ZEUS, bib:H1, bib:Gladilin} and from PYTHIA 6.4.21 simulations (green dotted line) and Statistical Hadronization Model~\cite{bib:SHM} (gray line) are shown.}
\end{figure}

As an example, the left panel of Fig.~\ref{fig:DCross} shows the cross section for prompt D$^0$ production measured by ALICE in pp collisions at $\sqrt{s} = 7$ TeV~\cite{bib:Dmes_pp_7}, compared with predictions from perturbative QCD calculations (FONLL and GM-VFNS). The right panel of Fig.~\ref{fig:DCross} shows the ratios of $p_{_{\rm T}}$-integrated cross sections for D$^0$, D$^+$, D$^{\ast +}$ and D$^+_s$ production in pp collisions at $\sqrt{s} = 7$ TeV~\cite{bib:Ds_pp_7}, compared with results from other experiments, PYTHIA simulations and SHM predictions. Cross sections for all the D mesons show a good compatibility with FONLL and GM-VFNS theoretical predictions within the uncertainties; their ratios are also in good agreement with the other measurements and with predictions within the uncertainties.

\section{D-hadron correlation: analysis strategy and results}
The analysis has been performed on a minimum bias sample of $3.14 \cdot 10^8$ pp events at $\sqrt{s} = 7$ TeV and on a minimum bias sample of $1.3 \cdot 10^8$ p-Pb events at $\sqrt{s_{\rm NN}} = 5.02$ TeV.
D-meson candidates (\textit{trigger} particles) are correlated with charged hadrons (\textit{associated tracks}) with $|\eta| < 0.8$ and the difference in pseudorapidity $\Delta\eta$ and azimuthal angle $\Delta\phi$ is evaluated. The contribution of background candidates is subtracted using the correlation distribution of candidates in the sidebands of the invariant mass distribution, normalized by the ratio of background candidates in signal and sideband regions. Corrections are applied to account for:
\begin{itemize}
\item Associated track reconstruction efficiency. Each D-hadron correlation is weighted by the inverse of the track reconstruction efficiency of the associated track, computed from a MC simulation as a function of the track $p_{_{\rm T}}$ and $\eta$ and the $z$ coordinate of the primary vertex;
\item D-meson selection efficiency. Each correlation entry is weighted by the D-meson reconstruction and selection efficiency, evaluated as a function of $p_{_{\rm T}}$ and the event multiplicity from a MC simulation;
\item Limited detector acceptance and detector spatial inhomogeneities. A correction factor is obtained by correlating D mesons from an event with tracks from other events (\textit{Event mixing}) with similar $z$ of the primary vertex and multiplicity and normalizing the correlation distribution obtained to its value in $(\Delta\phi$,$\Delta\eta) = (0$,$0)$;
\item Beauty feed-down. The fraction of D mesons coming from B meson decays is evaluated using the reconstruction efficiencies of prompt and secondary D mesons and the $p_{_{\rm T}}$-differential cross-section of the latter, obtained using FONLL calculations and the EvtGen package, as described in~\cite{bib:Dmes_pp_7}. MC simulations based on PYTHIA are used to obtain a template of angular correlations between D mesons from B meson decays and charged hadrons, which is then subtracted from the inclusive D-hadron correlation distribution obtained from data.
\end{itemize}

\begin{figure}[!ht]
\begin{minipage}{0.48\linewidth}
\includegraphics[width=\linewidth]{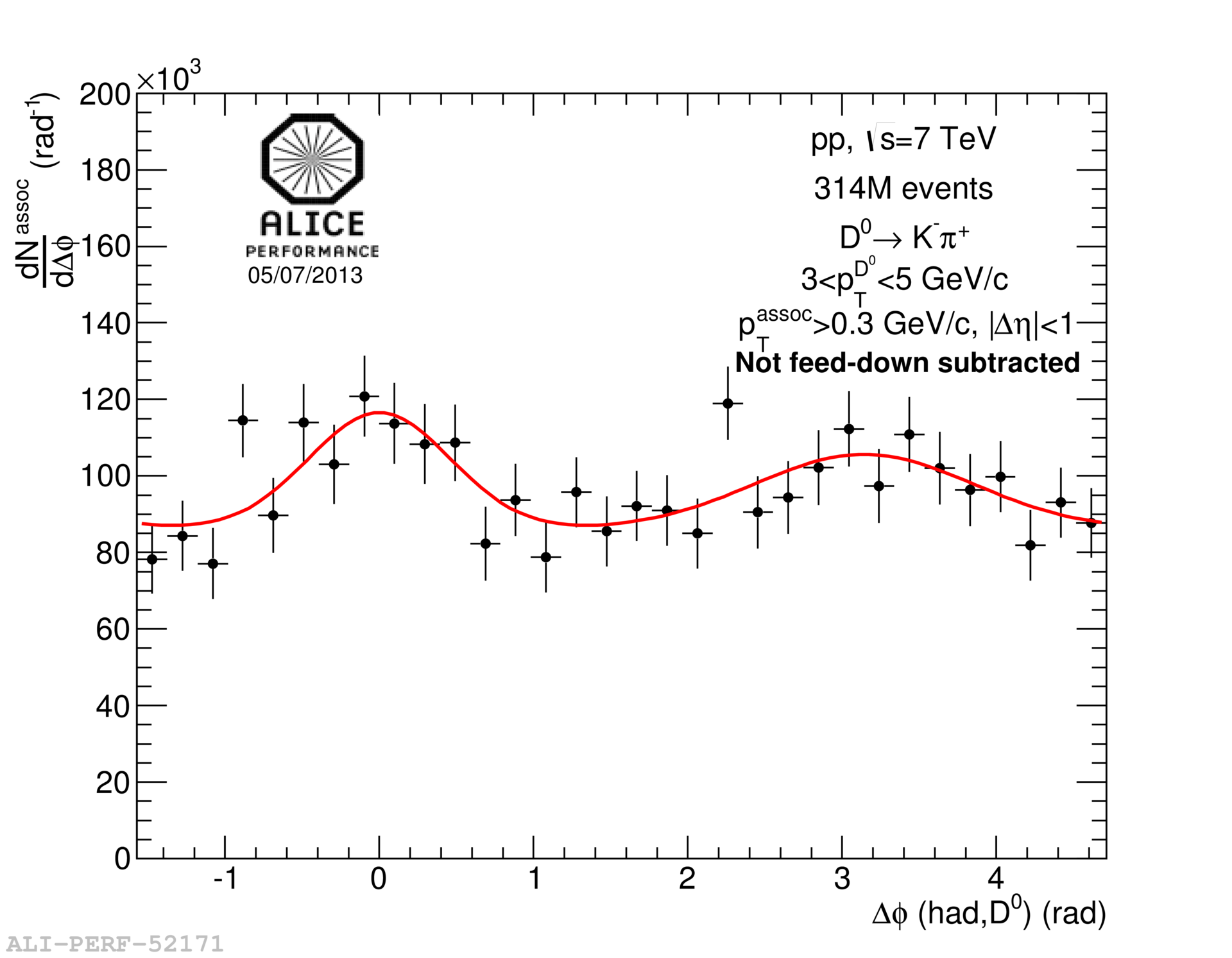}
\end{minipage}
\hspace{0.6pc}
\vspace{-0.5pc}
\begin{minipage}{0.48\linewidth}
\includegraphics[width=\linewidth]{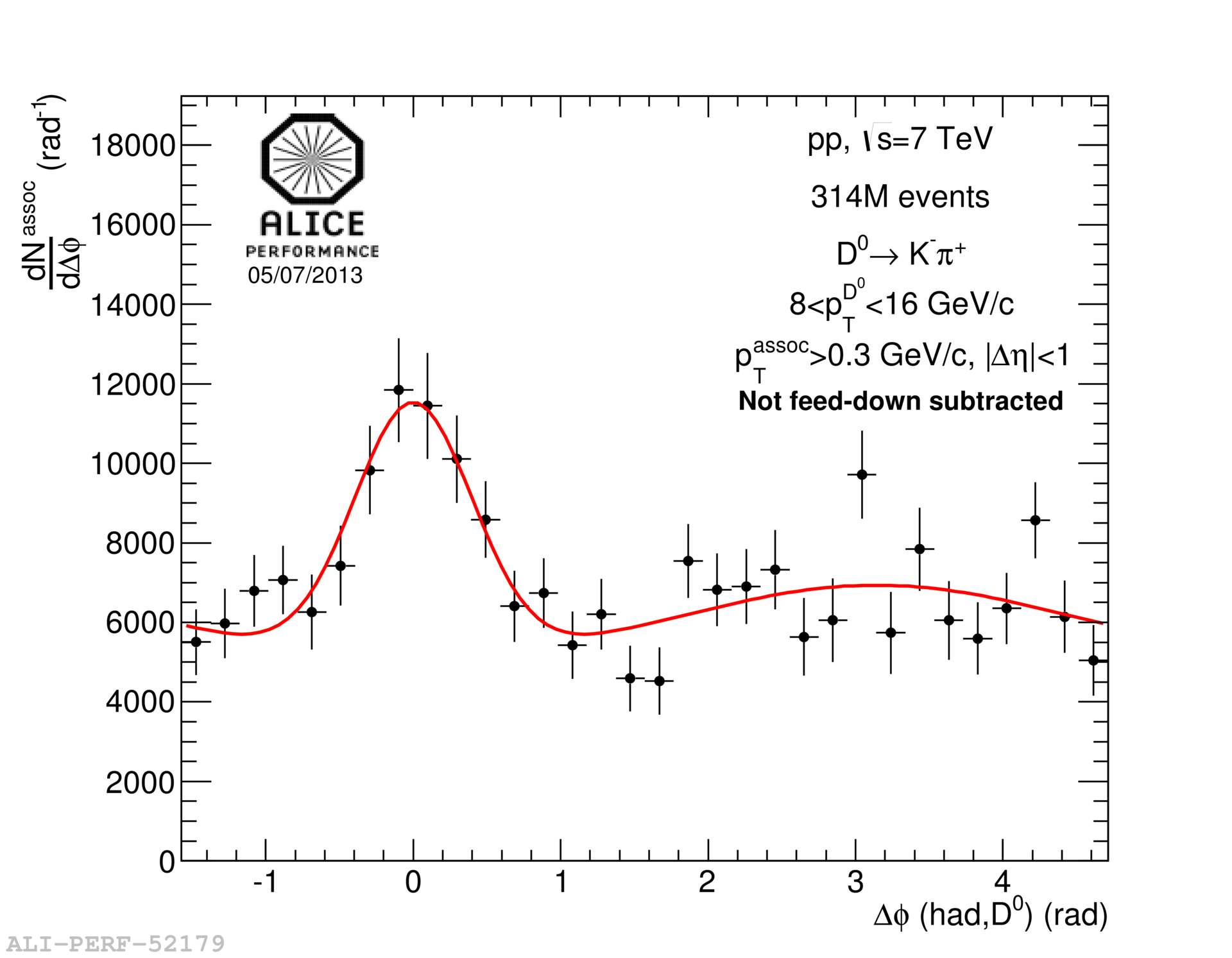}
\end{minipage}
\caption{\label{fig:D0_pp}D$^0$-hadron azimuthal correlations for $3 < p_{_{\rm T}}({\rm D}^0) < 5$ GeV/$c$ (left) and $8 < p_{_{\rm T}}({\rm D}^0) < 16$ (right) for associated tracks with $p_{_{\rm T}}^{h} > 0.3$ GeV/$c$ in pp collisions at $\sqrt{s} = 7$ TeV.}
\end{figure}

\begin{figure}[!ht]
\begin{minipage}{0.48\linewidth}
\includegraphics[width=\linewidth]{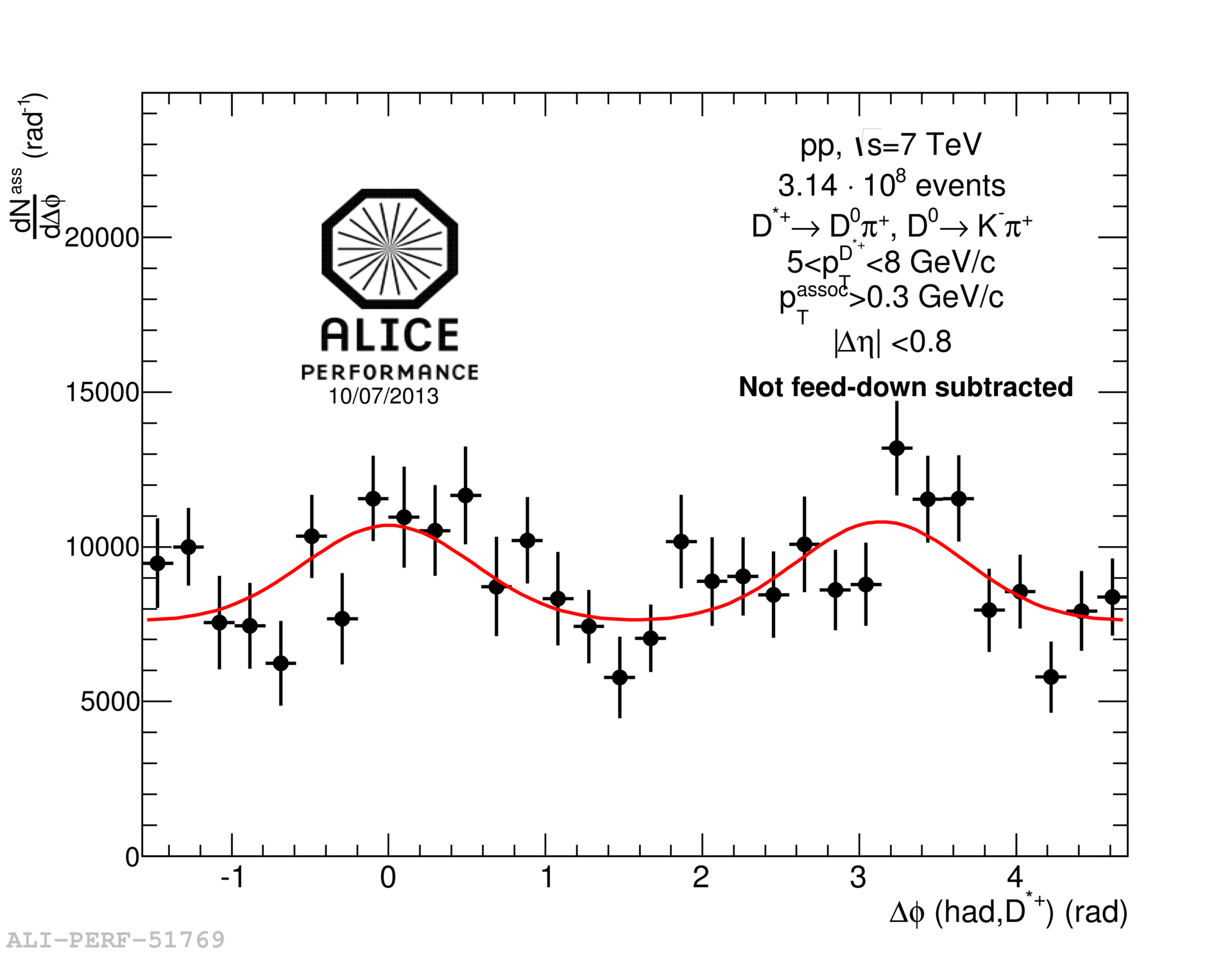}
\caption{\label{fig:D*_pp}D$^{\ast +}$-hadron azimuthal correlations for $5 < p_{_{\rm T}}({\rm D}^{\ast +}) < 8$ GeV/$c$ and associated tracks with $p_{_{\rm T}}^{h} > 0.3$ GeV/$c$, in pp collisions at $\sqrt{s} = 7$ TeV.}
\end{minipage}
\hspace{0.6pc}
\begin{minipage}{0.52\linewidth}
\vspace{-0.4pc}
\includegraphics[width=\linewidth]{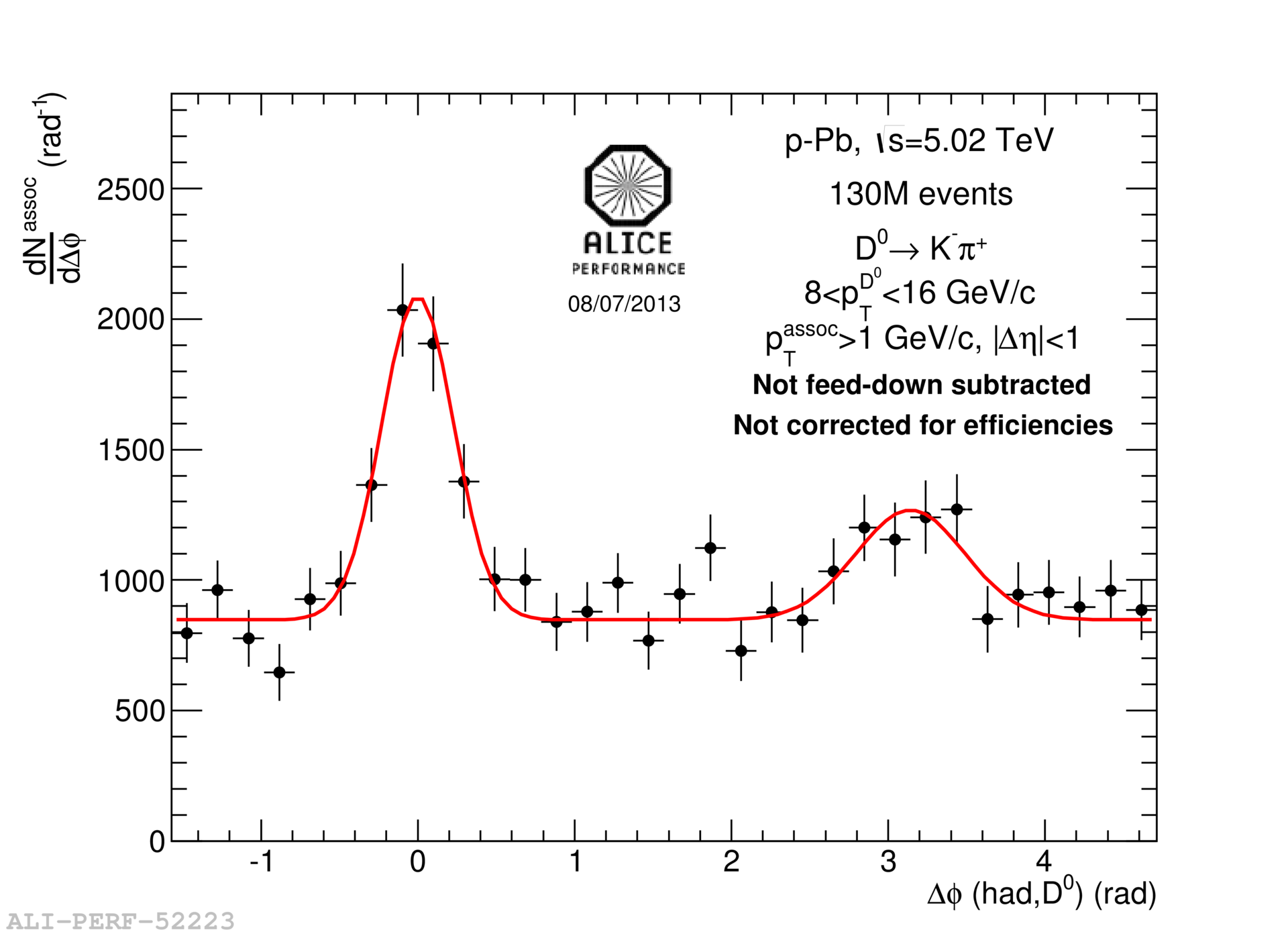}
\caption{\label{fig:D0_pPb}D$^0$-hadron azimuthal correlations for $8 < p_{_{\rm T}}({\rm D}^0) < 16$ GeV/$c$ and associated tracks with $p_{_{\rm T}}^{h} > 1$ GeV/$c$, in p-Pb collisions at $\sqrt{s_{\rm NN}} = 5.02$ TeV.}
\end{minipage}
\end{figure}

In the current analysis, the azimuthal correlations are not normalized to the number of D mesons and are not subtracted of the feed-down component.
After performing the steps listed above, 2-dimensional correlation plots ($\Delta\eta$,$\Delta\phi$) are obtained.
Due to the limited statistics available, the 2D correlations are projected on the $\Delta\phi$ axis, producing azimuthal correlation plots.
Figure~\ref{fig:D0_pp} shows D$^0$-hadron correlations for $3 < p_{_{\rm T}}({\rm D}^0) < 5$ GeV/$c$ (left) and $8 < p_{_{\rm T}}({\rm D}^0) < 16$ GeV/$c$ (right) in pp collisions at $\sqrt{s} = 7$ TeV, with associated tracks with $p_{_{\rm T}}^{h} > 0.3$ GeV/$c$. Near ($\Delta\phi \sim 0$) and away side ($\Delta\phi \sim \pi$) peaks are visible, though with significant fluctuations induced by the limited statistics available. In Fig.~\ref{fig:D*_pp} D$^{\ast +}$-hadron correlations are shown, for $5< p_{_{\rm T}}({\rm D}^\ast) < 8$ GeV/$c$ and a $p_{_{\rm T}}^{h}$ threshold of 0.3 GeV/$c$. In all correlations a fit is superimposed to guide the eye; it consists of two gaussians (with mean fixed at 0 and $\pi$) plus a constant baseline evaluated from points around $\pm \pi/2$ (using Zero Yield At Minimum assumption).

The feasibility of the analysis in p-Pb collisions at $\sqrt{s_{\rm NN}} = 5.02$ TeV and Pb-Pb collisions at $\sqrt{s_{\rm NN}} = 2.76$ TeV with the currently available statistics is being studied.
Figure~\ref{fig:D0_pPb} shows D$^0$-hadron correlations for $8 < p_{_{\rm T}}({\rm D}^{\ast +}) < 16$ and associated tracks with $p_{_{\rm T}}^{h} > 1$ GeV/$c$ in p-Pb collisions at $\sqrt{s_{\rm NN}} = 5.02$ TeV, with superimposed the same fit function described above.

\section{Conclusions and outlook}
We have presented measurements of D meson-hadron angular correlations in pp collisions at $\sqrt{s} = 7$ TeV and in p-Pb collisions at $\sqrt{s_{\rm NN}} = 5.02$ TeV.

The extension of the analysis to Pb-Pb collisions is challenged by the very low D-meson signal/background ratio and the large amount of tracks uncorrelated to D mesons. Improvement in the performance is expected with the ALICE upgrade ($\sim$2018-2019)~\cite{bib:LoI}, where a huge increment of statistics ($\times 100$) and a striking increase of the S/B ratio for D-meson reconstruction, thanks to the improved resolution on the track impact parameter provided by the new ITS, are expected in Pb-Pb collisions. This will reduce the uncertainty on the background subtraction, one of the main limiting factor in the Pb-Pb analysis, allowing thus to study nuclear matter effects on D-hadron correlations by comparison with pp and p-Pb results.

\end{document}